\documentclass[aps,showpacs,two column]{revtex4}

\usepackage{graphicx}
\usepackage[squaren]{SIunits}

\begin{document}
\title{Origin of superconductivity in nominally ``undoped"
T'-La$_{2-x}$Y$_{x}$CuO$_{4}$ films}

\author{L. Zhao, R. H. Liu, G. Wu, G. Y. Wang, T. Wu, X. G. Luo and X. H. Chen}
\altaffiliation{Corresponding author} \email{chenxh@ustc.edu.cn}
\affiliation{Hefei  National Laboratory for Physical Science at
Microscale and Department of Physics, University of Science and
Technology of China, Hefei, Anhui 230026, People's Republic of
China}

\begin{abstract}
We have systematically studied the transport properties of the
La$_{2-x}$Y$_{x}$CuO$_{4}$(LYCO) films of T'-phase ($0.05\leq x \leq
0.30$). In this nominally ``undoped" system, superconductivity was
acquired in certain Y doping range ($0.10\leq x \leq 0.20$).
Measurements of resistivity, Hall coefficients in normal states and
resistive critical field ($H^\rho_{c2}$)in superconducting states of
the T'-LYCO films show the similar behavior as the known Ce-doped
n-type cuprate superconductors, indicating the intrinsic
electron-doping nature. The charge carriers are induced by oxygen
deficiency. Non-superconducting Y-doped Pr- or Nd-based T'-phase
cuprate films were also investigated for comparison, suggesting the
crucial role of the radii of A-site cations in the origin of
superconductivity in the nominally ``undoped" cuptates. Based on a
reasonable scenario in the microscopic reduction process, we put
forward a self-consistent interpretation of these experimental
observations.
\end{abstract}
\vskip 15 pt
\maketitle

\section{INTRODUCTION}

After 20 years of intense research since the discovery of
high-temperature superconductivity(HTSC)\cite{bednorz}, many
interesting physical phenomena unique to the cuprate superconductors
are better understood. Although the underlying mechanism for HTSC
remains elusive and unresolved, it has been widely accepted
 \cite{orenstein} that the parent compounds of all the known
superconducting cuprates are half-filled Mott antiferromagnetic
insulators, in which the strong electronic correlation exists. Upon
doping holes or electrons into CuO$_2$ sheets, the
antiferromagnetism is suppressed, and superconductivity can be
induced in a certain concentration range of charge carriers.

Recently, Tsukada et al discovered a new class of cuprate
superconducting films, T'-(La, R)$_{2}$CuO$_{4}$ (LRCO, R=Sm, Eu,
Gd, Tb, Lu, and Y), which are nominally ``non-doped" because
R$^{3+}$ is the isovalent cation with La$^{3+}$ and the substitution
can not provide net charge carriers to CuO$_2$ planes \cite{naito2}.
In particular, the La-based T'-phase cuprates are metastable and
have not be acquired by conventional bulk synthesis now. These
T'-phase LRCO films were synthesized by molecular beam epitaxy
(MBE), and T$_c$ above 20K could be achieved in certain R doping
ranges. the authors claimed that the electron doping via oxygen
deficiencies was, at least, not a main source of charge carriers and
T'-LRCO are most plausibly conventional half-filled ``band
superconductors" with weak electronic correlation \cite{naito3,
naito4}. Their viewpoints arose strong skepticism because it
contradicts the acknowledged basic picture of HTSC in last 20 years.

However, further study on the physical properties of T'-LRCO is
lacking because the sample preparation is quite difficult due to the
metastability of La-based T'-phase structure and the repeatability
far from satisfaction by other groups. For example, an unsuccessful
attempt was reported recently\cite{Idemoto}, indicating that the
superconductivity of films is very sensitive to preparation
conditions. By pulsed laser deposition(PLD), Greene and co-workers
also prepared T'-LYCO films for only x=0.15 with partial success
\cite{greenelyco}. At present, it is urgent to perform systematic
study on LRCO physical properties.

Here using dc magnetron sputtering, we have been able to fabricate
the superconducting La$_{2-x}$Y$_{x}$CuO$_{4}$ (LYCO) films of pure
c-axis oriented T'-structure with high repeatability
\cite{lzhaoapl}.  In this paper, we systematically investigated
in-plane electronic transport and Hall coefficients of T'- LYCO
films with different Y contents: $x=0.05, ~0.10, ~0.125, ~0.15,
~0.20$ and $0.30$. The superconducting phase diagram and effect of
oxygen content on transport properties are studied. For films with
$x=0.125$ and $0.20$, the magnetic field dependence of
superconducting transitions was also measured to get the resistive
critical field $H^\rho_{c2}(T)$.

All these results show the similar behavior as the known
electron-doped T'-214 R$_{2-x}$Ce$_{x}$CuO$_{4}$ (R=Pr, Nd, ...)
systems, suggesting the same electron-doping nature in LYCO. The
charge carriers in T'-LYCO are induced by oxygen deficiency during
the reduction process. For comparison, we also studied the isovalent
substitution in  Nd$_2$CuO$_4$(NCO) and Pr$_2$CuO$_4$(PCO) systems.
No superconductivity was found. Combining our data and the results
in references \cite{naito2, naito3, naito4}, we present an approving
and self-consistent interpretation on the origin of
superconductivity in ``undoped" T'-214 cuprates, based on a
reasonable scenario about the microscopic depletion process of
oxygen upon annealing.

\section{EXPERIMENT DETAILS}

La$^{3+}$ ion is the largest among lanthanide series. The detailed
analysis on the perovskite tolerance factor $t$ \cite{goodenough1,
goodenough2}, has told us that La-based 214 cuprates adopt the
T-type phase by high-temperature bulk processes, while T'-214 phase
prefers stabilizing only at low synthesis temperature. By
extrapolation of the T/T' phase boundary in the
La$_{2-y}$Nd$_y$CuO$_4$ system to $y=0$, it has been predicted that
undoped T'-La$_2$CuO$_4$ can be stable below 425$\celsius$
\cite{goodenough2}. The partial substitution of La$^{3+}$ by smaller
Y$^{3+}$ can reduce the average ion radius of A-site, and shift
slightly the T/T' phase boundary to higher temperature. But it is
still too low for bulk synthesis of La-based T'-phase. In comparison
with the solid-state reaction process, the preparation of thin films
can usually be realized at lower temperature. Furthermore the
appropriate epitaxy strain through the substrates may stabilize some
metastable phases \cite{Tsukada}. Therefore, we grew c-axis oriented
LYCO films of pure T'-phase  by dc magnetron sputtering. The
stoichiometric ceramic targets are prepared by conventional
solid-state reaction. Fig. 1 shows the x-ray diffraction(XRD)
pattern of a typical LYCO film with x=0.15, which is pure
T'-structure of c-axis orientation. Only peaks of ($0~0~2l$)
diffraction appear, and the c-axis lattice constant $c$ is around
12.44\AA. The typical full width at half maximum (FWHM) of rocking
curves at the (006) reflection is only 0.28\degree, which confirms
the good epitaxial growth of the T'-LYCO films.

\begin{figure}
\includegraphics[width=0.50\textwidth]{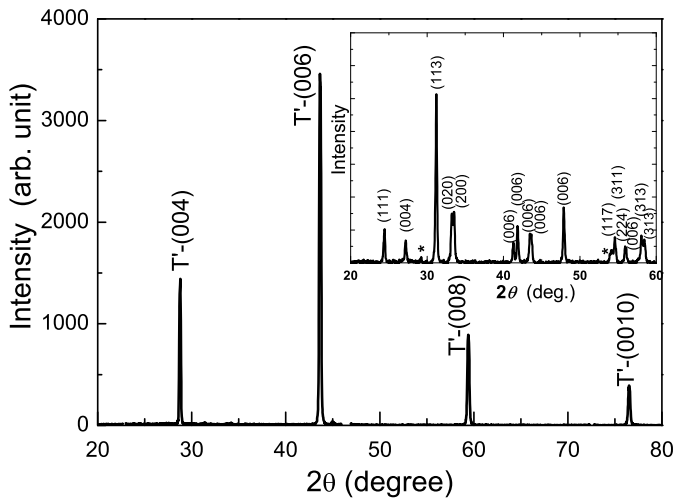}
\caption{ XRD pattern of La$_{1.85}$Y$_ {0.15}$CuO$_{4}$ thin film
deposited under T$_D$=690$\celsius$. the substrate reflections of
SrTiO$_3$ have been removed for clarity. The insert shows the XRD
data of corresponding target sintered at 1050$\celsius$ (the
un-indexed peak marked with stars is from a small quantity of
un-reacted Y$_2$O$_3$).\\}
\end{figure}

The XRD pattern of polycrystal LYCO target for $x=0.15$  is shown in
the insert of Fig. 1. Except the trace of un-reacted yttrium oxide,
all the peaks are well indexed assuming the La$_2$CuO$_4$-like
orthogonal T-214 structure. The calculated lattice constants of the
orthorhombic lattice are: a = 5.355\AA, b = 5.399\AA, and c =
13.136\AA, which are slightly less than those of undoped
La$_2$CuO$_4$, consistent with the partial replacement of larger
La$^{3+}$ (ion radii is 1.216\AA) by smaller Y{$^{3+}$ (1.075\AA).
There is a great contrast between T'-films and T-bulk synthesized at
different temperatures, consistent with above analysis about phase
stability.

The optimal growth conditions for x=0.15 LYCO films have been
discussed in detail elsewhere \cite{lzhaoapl}. Upon increasing Y
doping, the average  radii of A-site ions decrease, so does the $t$
factor, which leads to the shift of  T/T' phase boundary to higher
temperature, so that the corresponding optimal deposition
temperature T$_D$ is adjusted slightly to higher with increasing Y
doping. After deposition, all the films are annealed at
600$\celsius$ in high vacuum (near $10^{-4}$Pa) and finally followed
by very slow cooling to room temperature.

The Nd$_{2-x}$Y$_{x}$CuO$_{4}$ (NYCO) and Pr$_{2-x}$Y$_{x}$CuO$_{4}$
(PYCO) films were also prepared by dc sputtering for comparative
research, in which T'-214 structure is intrinsically stable. the
reduction process of films is the same as LYCO films.

The thickness of the typical films for transport measurements is
about 3000\AA~ . The measurements of resistivity and Hall
coefficients are carried out by standard 6-lead geometry using an
 ac resistance bridge(LR-700, Linear Research). The Ag electrodes are
made by Ag evaporation. Samples are mounted on the top of the
variable-temperature loading probe. Low temperature and high
magnetic field up to 14T were supplied by a Teslatron system (Oxford
Instruments)

\section{EXPERIMENTAL RESULTS}

\subsection{Superconductivity and Resistivity of T'-LYCO}

We first focus on the temperature-dependend resistivity $\rho(T)$ of
the T'-LYCO in zero magnetic field. Fig. 2 shows the the evolution
of $\rho(T)$ curves with different Y doping ($0.05\leq x \leq0.30$).

The film of x=0.05 (plotted in the insert of Fig. 2) shows a
insulating behavior in the whole temperature range like heavily
under-doped or oxygenated cuprates as NCCO\cite{wujiang, wangchprb}.

For all the $\rho(T)$ curves except for x=0.05, we observed metallic
behavior ($d\rho/dT>0$) in the high temperature range. In contrast
to the familiar linear temperature dependence in the hole-doped
cuprates like YBCO and LSCO, the nearly quadratic behavior of
$\rho$(T) in the normal state is observed (detailed analysis is
presented elsewhere \cite{lzhaoapl}). This kind of behavior is a
common feature in all the known electron-doped superconducting
cuprates near optimal and over doping, such as
Ln$_{2-x}$Ce$_{x}$CuO$_{4}$ (Ln=La, Pr, Nd, and Sm)\cite{Sawa,
Crusellas, Brinkmann, Greene}. It is usually considered as a quasi
2D Landau-Fermi liquid behavior due to electron-electron
scattering\cite{Tsuei}.

The metallic behaviors is maintained down to the resistivity minimum
temperature (T$_{min}$). Below T$_{min}$ , the resistivity shows a
upturn as in either hole- or electron- cuprates at low dopings. The
insulating upturn at low temperature are commonly interpreted in the
2D weak localization \cite{fournier} or the Kondo scattering by the
uncompensated Cu$^{2+}$ spins\cite{sekitani}.

\begin{figure}
\includegraphics[width=0.50\textwidth]{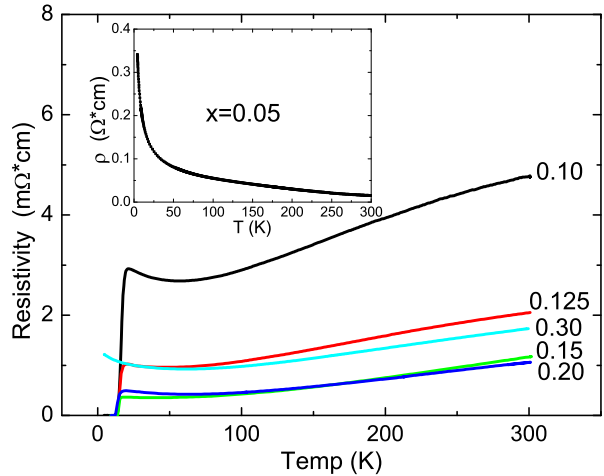}
\caption{ Resistivity vs. temperature for T'-LYCO films with various
Y contents: x=0.10, 0.125, 0.15, 0.20, and 0.30. The insert shows
the $\rho(T)$ of the film with x=0.05. The reduction processes for
all these films after deposition is the same.}
\end{figure}

\begin{figure}
\includegraphics[width=0.40\textwidth]{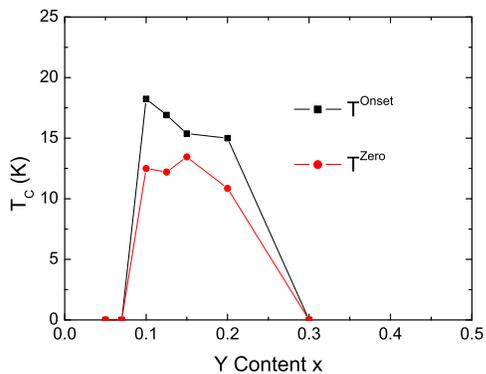}
\caption{ Phase diagram of T'-LYCO films as a function of Y content
x, determined from the resistivity data in Fig. 2. solid circles and
squares represent T$_c^{Zero}$ and T$_c^{Onset}$ respectively.}
\end{figure}

Although the Y$^{3+}$ ions doping can't bring net charge by
substitution of  La$^{3+}$, the resistivity of films reduces with
increasing Y content until to the minimum near x=0.20, suggesting
more electron doping by Y doping. As shown in the Fig.2,
superconductivity occurs when $0.10\leq x \leq 0.20$. But further
doping leads to the increase of $\rho$ and disappearance of
superconductivity. It is consistent with the evolution of Hall
coefficients (as discussed below). The superconducting phase diagram
for Y doping is plotted in Fig. 3. Superconductivity exists in
narrow Y doping range, i.e. $0.10\leq x \leq0.20$, which is in
agreement with previous results \cite{naito3}. The best T$_c^{Zero}$
is 13.5K near optimal doping x=0.15. T$_c^{Zero}$ does not vary much
with doping, differs from the usual superconducting ``dome" of other
cuprates in the diagram. One should note that the Y content is
indirectly relevant with the concentration of charge carriers. On
the contrary, heavy Y doping leads to larger resistivity and the
insulating behavior, instead of more metallic behavior induced by
overdoing in other cuprates.

\subsection{Resistive critical field $H^\rho_{c2}$}

Magnetic field dependence of superconducting transitions of two
T'-LYCO films (x=0.125, 0.20) is shown in Fig. 4 and 5. The magnetic
field was applied parallel to c-axis (shown in the main figures) or
parallel to ab plane (shown in the upper-right inserts of Fig. 4 and
5).

\begin{figure}
\includegraphics[width=0.50\textwidth]{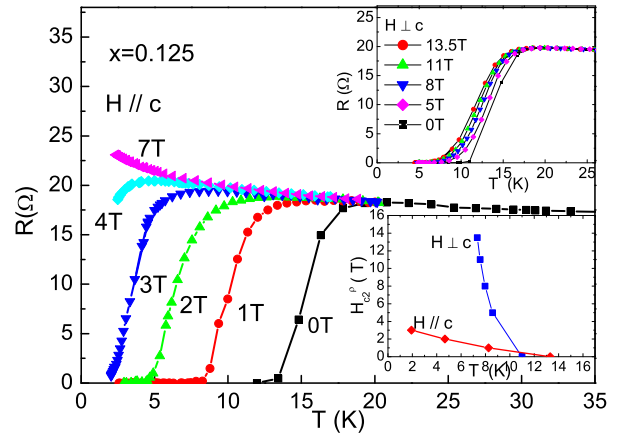}
\caption{ Resistive transition of T'-LYCO with x=0.125 at different
magnetic fields. the derived temperature dependence of $H^\rho_{c2}$
are shown in the bottom-right insert.}
\end{figure}

For $H // c$, the superconducting onset temperature (T$^{Onset}$)
for both films decreases and the width of the transition slightly
broadens at H=1T. Upon increasing the magnetic field, the transition
shifts to lower temperature and the $R(T)$ curves roughly parallel
to each other. no further broadening is found. The field of 7 Tesla
along c-axis is enough to kill the superconducting transition of
T'-LYCO film with x=0.125, but insufficient for x=0.20. The higher
upper critical field may be due to stronger vortex pinning by
in-plane oxygen deficiencies or local distortion of the lattice
induced by  Y substitution in fluorite layers in the T'- structure.

\begin{figure}
\includegraphics[width=0.50\textwidth]{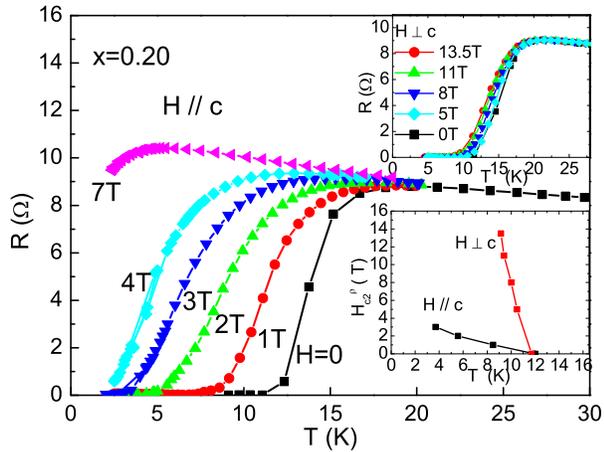}
\caption{ Resistive transition of T'-LYCO with x=0.20 at different
magnetic fields. the derived temperature dependence of $H^\rho_{c2}$
are shown in the bottom-right insert.}
\end{figure}

In Fig. 4 and 5, the most remarkable character is the strong
anisotropy in the effect of suppressing superconductivity by
magnetic field. When H// CuO$_2$ planes, fields up to 13.5 Tesla
just only slightly reduce Tc by a few Kelvins . This strong
anisotropy is a common feature in high-Tc cuprates due to their low
dimensionality and strong fluctuation effect.

From these $R(T)$ data in different magnetic fields, one can derive
a characteristic field which is often referred to as the resistive
upper critical field $H^\rho_{c2}$. The  choice of  criterion to
determine $H^\rho_{c2}$ remains with some uncertainty and
arbitrariness, because of the broadened flux-flow resistivity
curves. But as pointed in \cite{pfournier} and \cite{naitopg}, the
general trend of $H^\rho_{c2}$ is very insensitive to the criterion
used in electron-doped NCCO and PCCO. Here we adopt the zero
resistivity temperature $T^{Zero}_{C}$. Other criterions (10\% and
50\% of the transition), were tested, and also gave the similar
trend of($H^\rho_{c2}$) in the case of T'-LYCO.

$H^\rho_{c2}$ for H//c and H$\bot$c are shown in the bottom-right
inserts of Fig. 4 and 5 (the difference in Tc at H=0 between two
kinds of field configurations are from the remounting of samples on
the probe). For H$\bot$c, the derived values are too close to Tc
because of our experimental limits. It is difficult to deduce the
accurate trends towards low temperature. For H//c, the anomalous
positive curvature of $H^\rho_{c2}$(T) were observed in both of the
films. It seems to deviate from the conventional theory \cite{whh}.

The behavior of  $H^\rho_{c2}(T)$ in PCCO\cite{pfournier} and
NCCO\cite{naitopg} have been found to have the similar
temperature-dependence. It has been pointed out by Ong and
collaborators\cite{ong, yayuwang} that the resistivity is a bad
diagnostic to determine intrinsic $H_{c2}(T)$. The $H^\rho_{c2}(T)$
in cupates determined by resistivity measurements, can be actually
correlated with irreversibility line defined at the onset of flux
flow and near the crosser from the vortex solid to liquid in $H-T$
phase diagram in superconducting state. Only a very rough criterion
corresponding to a full recovery of the normal-state resistivity for
x=0.15 brings the characteristic field temperature dependence close
to the expected description by WHH theory with minus curvature and
saturate at low temperature. Because of the absence of pseudogap
above Tc in electron-doped cuprates, the Nernst measurements are
expected to serve as a much more accurate probe for $H_{c2}(T)$ in
T'-LYCO, as have been done in NCCO\cite{yayuwang}.

\subsection{Hall coefficients and origin of charge carriers in T'-LYCO}

At present. T'-LYCO films behave very similarly to the Ce-doped
n-type cuprates. No distinct difference is observed in above
experiments. To clarify the signature and concentration of charge
carriers in T'-LYCO, we examined the Hall coefficients of T'-LYCO
with different Y doping, as shown in Fig. 6.

\begin{figure}
\includegraphics[width=0.50\textwidth]{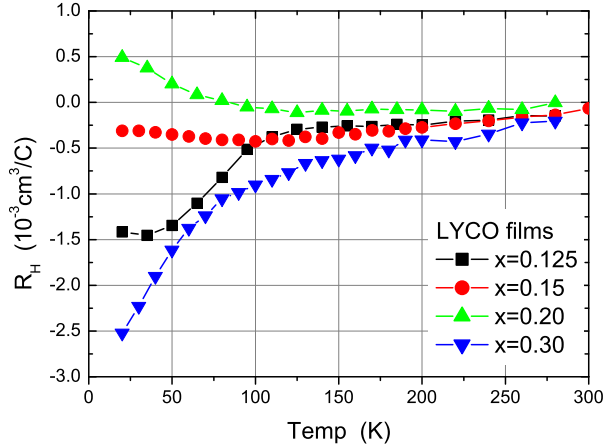}
\caption{ Temperature dependence of Hall coefficients of  T'-LYCO
films with different Y content x.}
\end{figure}

All the films show negative R$_{H}$ near room temperature. It
confirms the intrinsic electron-doped nature in T'-214 LYCO films.
For non-superconducting film with x=0.30, R$_{H}$ decreases upon
lowering temperature and becomes divergent approaching T=0. It is
typical behavior observe in lightly doped n-type cuprates, as NCCO
\cite{wangchprb}, PCCO\cite{Brinkmann} and LCCO \cite{Sawa}. The
R$_{H}$ of x=0.125 resembles the behavior of x=0.30, except the
upturn at low temperature and its smaller absolute value, suggesting
more concentration of electron carriers.

For x=0.15, R$_{H}$ decreases gradually with decreasing temperature
from room temperature down to about 100K. A slight upturn toward
zero occurs upon decreasing temperature further, which is analogical
to slightly underdoes NCCO or LCCO near the optimal doping in the
phase diagram. While for x=0.20, the upturn in low temperature is
observed, and the R$_{H}$ changes from negative in high temperature
to positive in low temperature. Such behavior is  similar to that in
electron-doped cuprates near optimal doping, which can been
considered  as a direct evidence of two types of charge carriers
\cite{wujiang}.

From above analysis on R$_{H}$, the variation of electron doping
n$_e$ with different Y contents is clear. n$_e|_{x=0.20}$ $>$
n$_e|_{x=0.15}$ $>$ n$_e|_{x=0.125}$ $>$n$_e|_{x=0.30}$, which is
consistent with previous resistivity data in Fig. 2.

Because Y$^{3+}$  and La$^{3+}$ are isovalent, no net charge can be
provided by Y doping. the most natural assumption is that the charge
carrier arise from oxygen deficiency since the superconductivity can
only be acquired by annealing films in high vacuum. But there is
still no way to determine the oxygen content in thin films
accurately and straightforwardly at present.

To confirm this assumption, we deposited LYCO(x=0.15) films at the
optimal temperature (T$_{D}$ =690$\celsius$), but annealed them in
lower vacuum (P$_{O_2}\sim 0.6\times 10^{-2}$Pa). These films are
still of pure T'-214 phase, but with less oxygen deficiency. Their
resistivity become much larger and insulating behavior are observed
at low temperature, as shown in the insert of Fig. 7. The
corresponding Hall coefficient R$_H$ was shown in Fig. 7, together
with R$_H$ of optimal T'-LYCO(x=0.15) for comparison. R$_H$ of
oxygenated films is negative in the whole temperature range and
decreases sharply with lowering temperature. The absolute value of
R$_H$ is much larger than that for optimal films, indicating much
less effective carrier density. It is consistent with its larger
$\rho$. Similar phenomena were observed during the oxygenation of
optimal NCCO \cite{wujiang} and PCCO\cite{Brinkmann}. It suggest
that oxygen deficiency is the source of electron doping in T'-LYCO
films.

\begin{figure}
\includegraphics[width=0.50\textwidth]{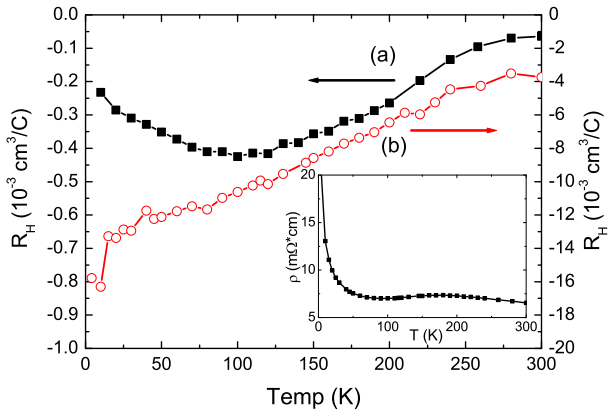}
\caption{ Temperature dependence of Hall coefficient R$_{H}$ for (a)
the optimal film annealed in high vacucum(near $10^{-4}$Pa), and (b)
the film annealed in lower vacuum(P$_{O_2}\sim 0.6\times
10^{-2}$Pa).  (a) and (b) are of the same Y doping x=0.15. The inset
shows the $\rho$-T data for the film (b).}
\end{figure}

\subsection{Comparative research on Y-doped Pr- and Nd-based T'-phases}

From above results, we can conclude that ``undoped"  T'-LYCO is
actually intrinsically electron doped and its charge carriers and
subsequent superconductivity in T'-LYCO film are induced by oxygen
deficiency. T'-LYCO seems to be an electron-doped counterpart of the
hole-doped superconductor La$_2$CuO$_{4+\delta}$ whose
superconductivity is induced by excess oxygenation.

Besides reducing \emph{t} factor to improve the stability of
T'-structure slightly, the role of Y in superconducting T'-LYCO is
still elusive. It is found that the superconductivity occurs only at
certain Y doping range($0.10\leq x \leq 0.20$). Does there exist
superconductivity in other ``undoped" T'-structure systems?  To find
 answer to these problems, comparative research on other T'-214
lanthanide cuprates is necessary.

It is known that in n-type superconducting cuprates the CuO$_2$
network is under tensile strain because of the bond-length mismatch
between the CuO$_2$ sheets and Ln$_2$O$_2$ fluorite layers
\cite{ytzhu, ytzhu2}. The tensile CuO$_2$ is apt to receive electron
doping and be reduced below the formal oxidation state
(CuO$_2$)$^{2-}$. Doping electrons into CuO$_2$ planes will increase
the Cu-O bond length and release the tension. Larger cations can
introduce more tensile strain. Naturally more stretch will increase
the capability to receive electron doping by CuO$_2$ sheets.
Countrawise, if the Cu-O bond length in the T' structures is too
small, as in Gd214 system(a$\sim $3.89\AA), the planes do not
readily accept election doping, no superconductivity is found with
Ce substitution \cite{smith}.

In Ref \cite{naito2}, the series of R$_2$CuO$_4$ films with
increasing radii of the R$^{3+}$ ions (R=Tb, Gd, Eu, Sm, Nd, Pr, La)
are fabricated by MBE. Their resistivity decreases gradually with
increasing ionic radius (shown in Fig. 3 in \cite{naito2}), the
trend is consistent with above analysis.

The Cu ions are reduced by in-plane deficiency and the stretch can
be partly released. Since the bond energies between Cu-O is less
than R-O \cite{bondingenergy}, It suggests more likely that main
oxygen deficit is mainly introduced at O(1) sites while O(2) sites
remain intact in the reduction. The in-plane oxygen defect created
by reduction have been evidenced in  infrared transmission, Raman
and ultrasonic studies in NCCO\cite{Raman, ultrasonic}and
PCCO\cite{infrared} systems .

In our work, we have prepared Nd$_{2-x}$Y$_{x}$CuO$_{4}$(NYCO) and
Pr$_{2-x}$Y$_{x}$CuO$_{4}$(PYCO) films with Y content x=0.15. The
temperature dependences of resistivity are shown in Fig. 8, in which
T'-214 structure is intrinsically stable even for x=0 and the role
of Y to stablize T'-structure is needlessly. It is convenient to
study the size effect at A-site cation in T'-phase. Furthermore, to
study the effect of cation disorder, the undoped
Pr$_{2}$CuO$_{4}$(PCO) films were also prepared for comparison. All
the films are of pure c-axis oriented T'-phase, and annealed in the
high vacuum.

\begin{figure}
\includegraphics[width=0.50\textwidth]{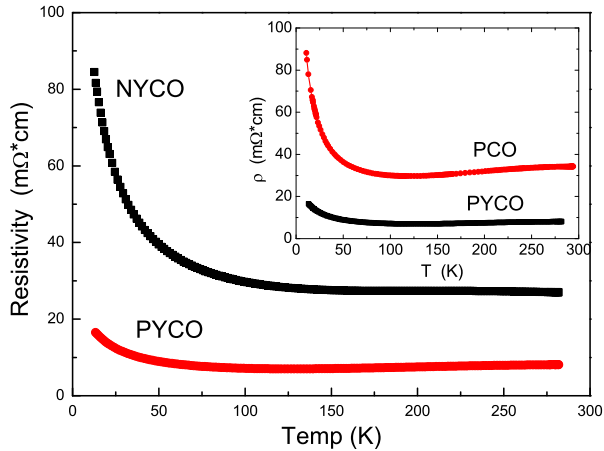}
\caption{Temperature dependence of resistivity of
Pr$_{1.85}$Y$_{0.15}$CuO$_{4}$(PYCO) and
Nd$_{1.85}$Y$_{0.15}$CuO$_{4}$(NYCO), Comparison of resistivity of
undoped Pr$_{2}$CuO$_{4}$(PCO) and PYCO is shown in the insert.}
\end{figure}

All films show insulating-like behavior and no superconductivity is
found, the resistivity of PYCO films is  smaller than of NYCO, which
is consistent with above analysis because of the larger average ion
radii at A-site in PYCO. In T'-phase, the larger cations at A-site,
the stronger driving force for the acceptance of electrons into
CuO$_2$ planes and thereafter the more electron doping by oxygen
vacancies are created during the reduction, as has been observed in
many early experiments on the preparation of T'-214 cuprate ceramics
\cite{kato, tao, ytzhu2}.

To find out why PYCO and NYCO are not superconducting in contrast to
LYCO, we measured their Hall coefficients from 280K down to 20K, as
shown in Fig. 9. The R$_H$ of superconducting LYCO films(x=0.15)  is
also plotted for comparison.

\begin{figure}
\includegraphics[width=0.50\textwidth]{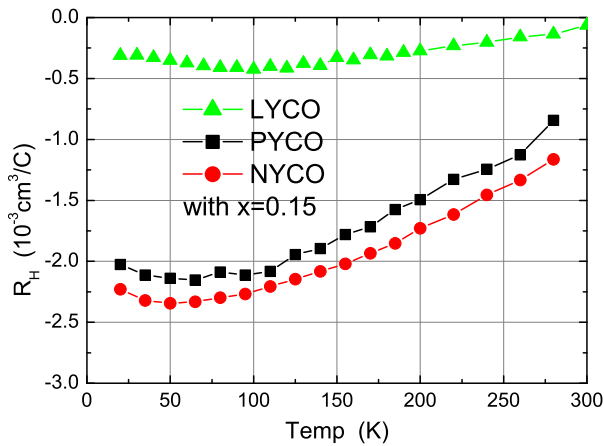}
\caption{Temperature dependence of Hall coefficients R$_{H}$ of
LYCO, PYCO and NYCO fims with the same Y content, x=0.15.}
\end{figure}

The  Hall coefficients R$_{H}$ of PYCO and NYCO films are both
negative in the whole temperature range and decrease gradually upon
lowering temperature with a slight upturn below 50K. The absolute
value of R$_H$ is much larger than that for LYCO films, indicating
much less of effective carrier density, consistent with resistivity
data. As pointed out in \cite{ytzhu}, the critical amount of
electron doping necessary to induce a transition from
antiferromagnetic insulator to superconductor is nearly fixed in
cuprates. The low carrier density should  be the main reason for the
absence of superconductivity in NYCO and PYCO.

For comparison, the resistivity of PYCO  and undoped PCO is shown in
the insert of Fig. 8. The partial substitution of Y for Pr greatly
reduces the resistivity, suggesting more oxygen vacancies induced by
Y doping. Although at present there is no the quantitative theoretic
calculation and direct experimental evidence about the oxygen
deficiency, we think it a reasonable microscopic scenarios that the
local distorted fluorite structure weaken the local binding of Ln-O
and make the further loss of oxygen in fluorite layer. Therefore, Y
doping leads to the additional oxygen deficiency,  and the increase
of the effective charge carriers, consequently.

\section{DISCUSSION}

We have shown the T'-LYCO is in nature electron-doped, like other
Ce-doped cuprate superconductors as NCCO and PCCO. The electron
doping is considered to origin from oxygen deficiency.

Although the strain effect by the substrates may help to stabilize
the T' structure to some degree, this kind of effect can be
negligible because the films we used in measurements are much
thicker than 1000\AA. In addition, two kinds of substrates,
SrTiO$_3$(STO) and
[(LaAlO$_3$)$_{0.3}$(Sr$_2$AlTaO$_6$)$_{0.7}$](LSAT) are used and no
apparent difference is found in the transport properties in our
experiments. Since we are working with thin films, the composition
inhomogeneity have also been ruled out.

The apical or interstitial oxygen problem, which is thought
deleterious and even deadly to superconductivity, are also excluded
in our discussion, because these films are annealed sufficiently in
the vacuum where the oxygen partial pressure is far below the level
in reducing single crystals or ceramics in flowing inert gas. The
removal of apical oxygen is complete and actual stoichiometric ratio
for oxygen should be less than 4.0, according to previous
determination of the oxygen content in bulk after sufficient
reduction\cite{ytzhu2}

Based on our present results and data from Tsukada  et al
\cite{naito2, naito3}, we present a most likely and reasonable
mechanism of the superconductive in T'-(La, RE)$_{2}$CuO$_{4}$. The
electron doping origins from two different microscopic processes of
oxygen deficiency, and each is indispensable to trigger
superconductivity.

\textbf{Firstly}, the large average radius of cations at A-site
should make the CuO$_2$ sheets in enough tensile, which become the
internal driving force for the acceptance of electrons into  CuO$_2$
sheets to release the stretch strain. Driven by the internal stress,
oxygen deficiency is produced in reduce process, mainly at the O(1)
position in the CuO$_2$ planes because of the lower binding energy
at the O(1) site than that at the O(2) site in La$_2$O$_2$ fluorite
layers. The larger cations at A-site will introduce more oxygen
vacancies which are responsible for more charge carriers and the
suppression of the long-range antiferromagnetism.

People have found in previous experiments that there are more oxygen
vacancies produced as the average ion radius of A-site increases in
T'-214 systems during a constant reduction process \cite{ytzhu,
tao}. It also consistent with the results that the resistivity
decreases gradually with increasing lanthanide ionic radius in
annealed undoped Ln$_2$CuO$_4$ (Ln=Tb, Gd, Sm, Eu,La, Pr, Nd) as
shown in Fig. 3 in Ref.\cite{naito2}.

Furthermore the difference from ion size effect also leads to
different additional doping by oxygen deficiency, and modify the
superconducting phase diagram as a function of Ce content x in known
Ce-doped n-type cuprates films (as shown in Fig. 1 in Ref
\cite{naito2}). The shift of the highest Tc towards lower Ce
concentrations is also observed as increasing the substitution of La
to PCCO\cite{koike} or NCCO\cite{tao,kato} bulk systems. This
evolution with average size of A-site cations is  consistent with
our above analysis.

Nevertheless, in undoped T'-214 cuprates the charge carriers induced
only by in-plane vacancies are not enough to drive the CuO$_2$ into
superconducting states. The additional oxygen deficiency in fluorite
layers is needed. But in undoped Ln$_2$CuO$_4$ , the binding energy
in Ln-O is always larger than that in in-plane Cu-O, so that O(2)
can be too hard to be removed before the excess in-plane oxygen
vacancies destabilize the T'-structure and lead to chemical
decomposition of samples during the further reduction.

\textbf{Secondly}, the partial substitution of smaller Y$^{3+}$ ions
for La$^{3+}$ leads to local strong distortion in the fluorite
Ln$_2$O$_2$ layers. It will be apt to further produce oxygen
vacancies during the reduction. The strong distortion in fluorite
layers by Y substitution is thought to decrease the local binding
energy to be overcome, and then creates additional oxygen vacancies
in fluorite layers in reduction. It has been demonstrated in the
comparison of doped and undoped PCO films.

Therefore the total oxygen deficiency, created in two ways proposed
above, will be adequate to induce superconductivity in T'-LRCO. As
proposed by Zhu and Manthiram \cite{ytzhu}, the  transition from
antiferromagnetic insulator to superconductor occurs at the critical
amount of electron concentration, $n_c$, which is nearly fixed in
n-type cuprates. Although the Madelung energy or charge transfer gap
$\Delta$ decreases upon increasing ion radius at A-site, its effect
on $n_c$ is negligible.

Increasing Y doping will introduce more charge carriers. However,
excess Y doping induces the decrease of the average size of Ln at
A-site, which will hinder the oxygen depletion at O(1)-site.
Furthermore, the disorder induced by Y doping localizes the charge
carriers. Therefore excess Y will kill the superconductivity in
extreme case. The increasing resistivity and insulating upturn in
the T'-LYCO films with high doping(x=0.30), are observed in our
experiments.

Our  analysis above also naturally explains other results of Tsukada
and cooperators \cite{naito2}. Superconductivity has been found in
many ``non-doped" T'-(La, RE)$_{2}$CuO$_{4}$ cuprates as RE=Sm, Eu,
Gd, Tb, Lu, and Y, but neither Pr nor Nd. Pr and Nd are more
adjacent to La in lanthanide series, the difference of ion radii is
smaller than other lanthanide. It suggests that the local distortion
caused by Pr or Nd substitution is not as remarkable as by Y etc.
The fluorite structure remains almost intact during the usual
annealing process. Oxygen vacancies mainly exist in CuO$_2$ network,
and they can't afford necessary doping to trigger superconductivity
unless there is extra charge induce by aliovalent Ce substitution in
charge reservoir layers as in the (La,Nd,Ce)$_2$CuO$_4$ systems
\cite{tao}.

Our present results provide a more comprehensive understanding of
this new class of cuprates as well as the microscopic process of
oxygen reduction which is indispensable to acquire superconductivity
in electron-doped cuprates. Because of the metastability of La-based
T'-214 structure, people can only obtain superconducting T'-LRCO in
the form of thin films so far. Therefore, it is difficult and even
impossible to determine directly the oxygen occupancy at O(1) and
O(2) sites in films within present technologies such as conventional
X-ray or neutron diffraction. Bulk T'-La$_2$CuO$_{4}$ \cite{chou}and
even supercooling La$_{2-x}$Ce$_{x}$CuO$_{4}$ \cite{yamada} have
been synthesized in special precursor route. It indicates that the
bulk synthesis of superconducting T'-LYCO seems to be possible. Such
attempts is underway. Due to low sintering temperature, it will be
very difficult to achieved the good crystallization and homogeneity
in bulk. For the second processes of oxygen reduction we prepose
above, other experimental methods such as Raman, infrared
transmission and Extended x-ray absorption fine structure (EXAFS)
x-ray absorption near edge structure (XANES) etc is expected to
detect the distortion of the local atomic structure and thereafter
its effect on oxygen depletion. Corresponding studies are expected
in future study and are beyond our present research.

\section{CONCLUSION}

In summary, we systematically  investigated the transport properties
of newly discovered nominal ``non-doped" La$_{2-x}$Y$_{x}$CuO$_{4}$
($0.05\leq x \leq 0.30$). Detailed studies on the resistivity and
Hall coefficient of T'-LYCO films, strong anisotropic
H$^{\rho}_{c2}$(T) and its positive curvature, suggest the
similarity to the known cation-doped election-type cuprates
superconductors as NCCO or PCCO. This kind of so-called ``non-doped"
superconductor is intrinsically electron-doped and the effect of
oxygen content indicates that the charge carriers mainly arise from
oxygen deficiency obtained during the reduction process. The
fundamental picture \cite{anderson} depicted since the discover of
HTSC still holds at present. Further studies on the effect of the
in-plane partial substitution of magnetic or non-magnetic impurities
are in process.

Comparative experiments on non-superconducting Y-doped Pr- and Nd-
based T'-214 films exhibit the the crucial role of cation radius
played in the oxygen depletion during the reduction. The cooperation
of large-sized La$^{3+}$ ions with smaller Y$^{3+}$ enable the
sufficient oxygen deficiency in T'-structure to induce
superconductivity in T'-LYCO. We propose a reasonable scenario about
the microscopic reduction process. The large La$^{3+}$ ions at
A-site keep the CuO$_2$ in adequate tensile strain which enhances
in-plane oxygen deficiency during reduction. And the appropriate
local distortion induced by  the substitution of small Y$^{3+}$ in
fluorite layers lowers the binding energy to be overcome in the
oxygen depletion in fluorite layers. They are two indispensable keys
to induce enough electron doping to trigger superconductivity in the
nominally ``undoped" cuprates.

The work is supported by the Nature Science Foundation of China and
by the Ministry of Science and Technology of China (973 project No:
2006CB601001 and 2006CB0L1205), and by the Knowledge Innovation
Project of Chinese Academy of Sciences.

\clearpage

\end{document}